# Algorithms of an optimal integer tree labeling

Alexander Bolshoy[1,2] and Valery M. Kirzhner[2,3]

[1]Department of Evolutionary and Environmental Biology, University of Haifa, Haifa, Israel

[2]Institute of Evolution, University of Haifa, Haifa, Israel

[3]The Tauber Bioinformatics Research Center at University of Haifa

**Abstract**

Suppose we label the vertices of a tree by positive integers. The *weight* of an edge is defined by a monotonically increasing function of the absolute value of the difference of the labels of its endpoints. We define the *total cost* of the labeling to be the sum of weights of all the edges.

The problem we consider is that of determining for a given tree G and given a labeling of the leaves of G the minimum total cost labelings of G. In this paper we present an algorithm that works for any cost function satisfies the condition of monotony mentioned above. In a case of the function defined as the absolute value of the difference of the labels the fast algorithm is presented.

## 1. Introduction

In the graph theory vertex labeling related problems were intensively studied [1]. Typically, the problems can be described as follows: for a given graph, find the optimal way of labeling the vertices with **distinct** integers. If we want to find the labeling which minimizes the total 'length' sum of the edges, we have the minimum sum problem. In a similar way the so-called bandwidth and cutwidth problems are defined. The problems and their solutions were described in [1-3, 5, 6]. These problems came up in connection with applications in network addressing, code design or similar IT problems. The problems presented in this paper are motivated by molecular evolution, and tree node labels are connected with lengths of modern and ancestral proteins.

In this paper we give two algorithms: a pseudo-polynomial algorithm to solve the max sum problem on trees for a monotonically increasing cost function θ, and a linear algorithm to solve max sum problem on binary trees for the Manhattan cost function θ: $|\pi(v) - \pi(w)|$. The first algorithm uses dynamic programming techniques; the second algorithm uses the properties of the Manhattan distance.

## 2. Preliminaries

Let *G* be a tree with *n* leaf nodes, vertex set *V(G)* and edge set *E(G)*. N=|V(G)|. Let us number the leaf nodes of G: 1, 2, …, n. Let us number the root of G: N. An **integer labeling π** of G is a mapping π from G to a set of positive integers. Let us denote integer labeling of the leaf nodes of G ($\pi(1) = p_1$, …, $\pi(n) = p_n$}. Let us denote by $g_{min}$ and $g_{max}$ min and max integers labeling leaf nodes: $g_{min}=\min\{p_i\}$, $g_{max}=\max\{p_i\}$; $m=g_{max}-g_{min}+1$

The θ-sum of a numbering π is $S(G) = \sum_{vw \in E(G)} \theta(|\pi(v) - \pi(w)|)$, where θ(x) is a monotonically increasing function of nonnegative *x* θ(0)=0. An example of such a function is $|\pi(v) - \pi(w)|^\lambda$. In case of λ=1 we obtain the Manhattan distance.
The **Manhattan sum** of a numbering π is $S(G) = \sum_{vw \in E(G)} |\pi(v) - \pi(w)|$

### 2.1. Arbitrary tree and an arbitrary cost function

Given a tree G, an integer labeling of the leaves of G (p1, …, pn) and a monotonically increasing cost function θ, the minimum sum problem is to find a labeling which minimizes the total cost: $(G) = \sum_{\forall\{vw\} \in E(G)} \theta(|\pi(v) - \pi(w)|)$ over all π.

### 2.2. A binary tree problem

Given a binary tree G, an integer labeling of the leaves of G ($p_1$, …, $p_n$) and the Manhattan cost function θ: $|\pi(v) - \pi(w)|$ the minimum sum problem is to find the labelings which minimize $s(G) = \sum_{\forall\{vw\} \in E(G)} |\pi(v) - \pi(w)|$ over all π.

## 3. Problem solutions
### 3.1. DP algorithm (for the problem 2.1)
#### 3.1.1. DP algorithm for a binary tree

*Up phase. A procedure called DP_up calculates the costs $S_k(i)$ of all nodes V(G) of the tree G, given a cost function θ.*

Let us suppose we have proven that all labels of the optimal labeling must be in the interval [$g_{min}$, $g_{max}$]. (For the proof, see Lemma 1 below.) Let us compute, for all integers of [$g_{min}$, $g_{max}$], for each node *k* in the tree, a quantity $S_k(i)$. This quantity will be interpreted as the minimal cost, given that node *k* is assigned integer *i*, to the subtree with the node *k* as a root in the tree. In other words, it is the minimal cost of stretches in the subtree, which starts at node *k*. It should be immediately apparent that if we can compute this for all nodes, we can compute it for the root in the tree, in particular. If we can compute it for the root node (the index of the root is N), then we can simply choose the minimum of these values:

$$S = \min_i S_N(i) \tag{3.1}$$

Given labeling of the leaf nodes of G ($p_1, \ldots, p_n$) at the tips of the tree the S(i) are easy to compute. The cost is 0 if the observed integer $p_k$ is integer *i*, and infinite otherwise.

$$S_i(j) = \begin{cases} 0 & \text{if } j=p_i \\ \infty & \text{otherwise} \end{cases}$$

Now all we need is an algorithm to calculate the $S_a(i)$ for the immediate common ancestor of two nodes. This is very easy to do. Let us show a proof for a binary tree and after that for a common case. Suppose that the two descendant nodes of the node *a* are called *l* and *r* (we distinguish between a number of a node and its label). For the immediate common ancestor of the nodes *l* and *r*, node a, we need only compute

$$S_a(i) = \min_j[\theta(|i-j|) + S_l(j)] + \min_k[\theta(|i-k|) + S_r(k)], \forall i,j \in [g_{min}, g_{max}] \tag{3.2}$$

The interpretation of this equation is immediate. The smallest possible cost given that node a is assigned i is the cost $\theta(|i-j|)$ of the edge from the node a to the node l, plus the cost $S_l(j)$ of the left descendant subtree given that node l is in state j. We select that value of j which minimizes that sum. We do the same calculation for the right descendant subtree, which gives us the second term of (2.2). The sum of these two minima is the smallest possible cost for the subtree above node a, given that node a is in state i. This equation is applied successively to each node in the tree, doing a postorder tree traversal. Finally it computes all the $S_N(i)$, and then (2.1) is used to find the minimum cost for the whole tree. The complexity of the algorithm is

$$O(N*(g_{max}-g_{min})*(g_{max}-g_{min})) \tag{3.3}$$

***Down phase***. *A procedure named DP_down calculates the labels S(p) of all nodes p of the tree G, given the root N of G, given a cost function θ, and the costs $S_k(i)$ of all nodes V(G) of the tree G as calculated by DP_Up described above.*

Choose any integer *i* provides the minimum of the $S_N(i)$ - it is the root label. Doing a preorder tree traversal, label successively each inner node in the tree: for any inner node *p*, and given that a label *i* was reconstructed at its parent, the label *j* to be chosen is that for which $\theta(|i - j|) + S_j(p))$ is minimized. Notice that for inner nodes this state does not necessarily correspond with that for which $S_j(p)$ is minimum.

### 3.1.2. DP algorithm for an arbitrary tree.

***Up phase***. *A procedure DP_up calculates the costs $S_k(i)$ of all nodes V(G) of the tree G, given a cost function $\theta$.*

The common case is presented in a similar way. Suppose that the $k_a$ descendant nodes of the node *a* are called $b_j$. So,

$$S_a(i) = \sum_j \left[\min_{j1}\left(\theta(|i - j1|) + S_{j1}(b_j)\right)\right], \forall i,j \in [g_{min}, g_{max}], 2 \leq j1 \leq k_a \quad (3.4).$$

This equation is applied successively to each node in the tree, doing a postorder tree traversal. Finally it computes all the $S_N(i)$, and then (2.1) is used to find the minimum cost for the whole tree.

***Down phase***. As Up-phase above for 3.1.1.

*Comment 1*. This algorithm is a dynamic programming (DP) algorithm, because it solves the problem of finding the minimum cost by first solving some smaller problems and then constructing the solution to the larger problem out of these, in such a way that it can be proven that the solution to the larger problem is correct. An example of a dynamic programming algorithm is the well-known pairwise sequence alignment algorithm [7, 8, 11, 12]. Another example is Sankoff parsimony algorithm [9], which determines the minimum number of changes required in a given phylogeny when a cost is associated to transitions between character states.

Lemma 1.
All labels of the optimal integer labeling, which minimizes sum $\sum_{vw \in E(G)} \theta|\pi(v) - \pi(w)|$, are in the interval [$g_{min}$, $g_{max}$].
Proof.
Let us denote a set of vertices X=[$x_1$,…,$x_k$] $\pi(x_i) < g_{min}$. Let us introduce a new numbering $\pi'$: let us substitute labeling of all $x_i$: $\pi'(x_i) = g_{min}$. $\forall i, 0 \leq i \leq k \; \forall v \langle x_i, v \rangle \in E(G)$. It is easy to see that if k>0 then $S_{\pi'}(G) < S_\pi(G)$. It means that for optimal integer labeling $\pi$ the following is correct: $\forall v \in V(G) \pi(v) \geq g_{min}$. Likewise, $\forall v \in V(G) \; \pi(v) \leq g_{max}$. Quod erat demonstrandum.

### 3.2. Linear algorithm for a Manhattan sum for a binary tree

*Bottom-up stage.*

Doing a postorder tree traversal assign successively to each node in the tree an interval, according to the following rule: a parent interval is either an intersection of the intervals of its children or an interval that lies between these intervals in case that their intersection is empty.

*Top-down stage.*

Choose any integer from the interval assigned to the root node - it is the root label. Doing a preorder tree traversal, label successively each node in the tree by an integer from the interval assigned to this node which is the nearest to its parent label: It may be either the value equal to the parent label or the boundary value of the interval assigned with the node.

The proof of the correctness of the algorithm that solves the problem of finding an optimal integer labeling, which minimizes sum $\sum_{vw \in E(G)}|\pi(v) - \pi(w)|$, follows from the following lemmas.

Lemma 2. Root Optimal Label.
<u>A root label of the optimal integer labeling is in between the two descendant labels.</u>
Suppose that the root node is called *rt* (rt=N), suppose that its children are called *l* and *r*. If $(\pi(l) \leq \pi(r))$ then $\pi(rt)$ is in the interval $[\pi(l), \pi(r)]$ else $\pi(rt)$ is in the interval $[\pi(r), \pi(l)]$

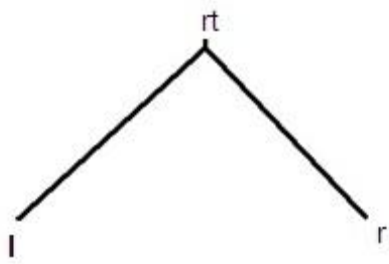

**Figure 1. If $(\pi(l) \leq \pi(r))$ then $\pi(l) \leq \pi(root) \leq \pi(r)$**

Proof.

Suppose that $\pi(rt) < \pi(l)$. Let us introduce a new numbering $\pi'$ by changing only the root label: $\pi'(rt) = \pi(l)$. It is easy to see that $S_{\pi'}(G) < S_\pi(G)$. It means that for optimal integer labeling $\pi$ the following is correct: $\pi(rt) \geq \pi(l)$. Likewise, we prove that for optimal integer labeling $\pi(rt) \leq \pi(r)$. Q.e.d.

Lemma 3. Root Optimal Interval.

All integers between the two children labels of the root label of the optimal integer labeling are equally suitable to label the root. Let us note $k = \pi(rt)$.

Proof. According to Lemma 2 if $(\pi(l) \leq \pi(r))$ then $\pi(l) \leq k \leq \pi(r)$. $\forall k \big(\pi(l) \leq k \leq \pi(r)\big) S_\pi(G) = k - \pi(l) + S_\pi(l) + \pi(r) - k + S_\pi(r)$, q.e.d.

Lemma 4. Leaf Parent Optimal Interval.

Every node of the optimal integer labeling that all its children are leaf nodes has a label between labels of its children.

Let $G$ be a binary tree with $n$ leaf nodes, vertex set $V(G)$ and edge set $E(G)$. Let us number the leaves of G: 1, 2, …, n. Let us denote integer labeling of the leaves of G $(\pi(1) = p_1, …, \pi(n) = p_n\}$.

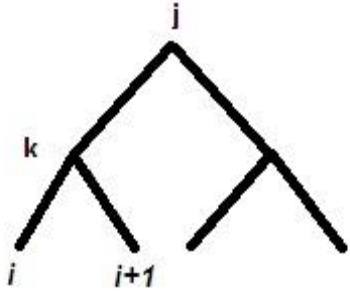

Figure 2. $\pi(i) \leq \pi(k) \leq \pi(i+1)$.

Suppose a vertex $k$ has the children $i$ and $i+1$ and the parent $j$. Suppose $p_i \leq p_{i+1}$. Let us prove that for optimal integer labeling $\pi(k) \geq p_i$. Suppose $\pi(k) < p_i$. Let us denote $(p_i - \pi(k)) = \delta$; $p_{i+1} - p_i = \gamma$. Let us introduce a new numbering π' by changing only the label k: $\pi'(k) = p_i$. Let us show that $S_{\pi'}(G) < S_\pi(G)$. Indeed, $S_{\pi'}(G) = S_\pi(G) - |\pi(k) - \pi(j)| - (p_i - \pi(k)) - (p_{i+1} - \pi(k)) + |\pi'(k) - \pi(j)| + (p_i - \pi'(k)) + (p_{i+1} - \pi'(k)) = S_\pi(G) + (|p_i - \delta - \pi(j)| - |p_i - \pi(j)|) - (p_i - \pi(k)) - (p_{i+1} - \pi(k)) + (p_i - p_i) + (p_{i+1} - p_i) = S_\pi(G) + (|p_i - \delta - \pi(j)| - |p_i - \pi(j)|) - \delta - (\delta + \gamma) - \gamma = S_\pi(G) - \delta$. Likewise, we prove that for optimal integer labeling $\pi(k) \leq p_{i+1}$. Q.e.d. $p_i \leq \pi(k) \leq p_{i+1}$.

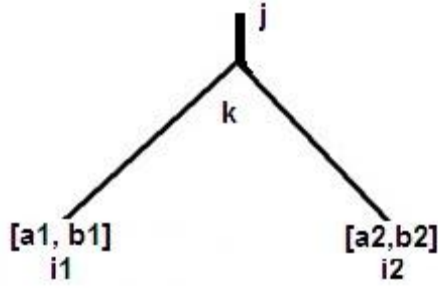

**Figure 3.** b1 ≤ π(k) ≤ a2.

Lemma 5. Parent Optimal Interval - I.

An optimal label of a parent lies between extreme values of optimal labels of its children. If an optimal integer labeling π provides the labels of two siblings $i_1$ and $i_2$ satisfying the conditions $a_1 \leq \pi(i_1) \leq b_1$ & $a_2 \leq \pi(i_2) \leq b_2$ then the label of their parent $k$ satisfies $\min(a_1, b_1, a_2, b_2) \leq \pi(k) \leq \max(a_1, b_1, a_2, b_2)$. Proof as for Lemma 2.

Lemma 6. Parent Optimal Interval - II. An optimal label of a parent in case of the empty intersection of the optimal intervals of its children lies between these intervals. If an optimal integer labeling π provides the labels of two siblings $i_1$ and $i_2$ satisfying the conditions ($a_1 \leq \pi(i_1) \leq b_1$) & ($a_2 \leq \pi(i_2) \leq b_2$) then if ($b_1 \leq a_2$) then the label of their parent $k$ satisfies the condition $b_1 \leq \pi(k) \leq a_2$ else if ($b_2 \leq a_1$) then $b_2 \leq \pi(k) \leq a_1$.
Proof:

1) $b_1 \leq a_2$. Let us assume $\pi(k) < b_1$; $\pi(k) = b_1 - \alpha$. Then we introduce a new labeling π' by changing labels for three nodes: $\pi'(k) = b_1$; $\pi'(i_1) = b_1$; $\pi'(i_2) = a_2$.

$$S_\pi(G) - S_{\pi'}(G)$$
$$= (|\pi(j) - \pi(k)| - |\pi(j) - \pi'(k)|)$$
$$+ (|\pi(k) - \pi(i_1)| - |\pi'(k) - \pi'(i_1)|)$$
$$+ (|\pi(k) - \pi(i_2)| - |\pi'(k) - \pi'(i_2)|)$$

$$(|\pi(j) - \pi(k)| - |\pi(j) - \pi'(k)|) = |\pi(j) - b_1 + \alpha| - |\pi(j) - b_1| = \alpha$$

$$(|\pi(k) - \pi(i_1)| - |\pi'(k) - \pi'(i_1)|)$$
$$= \left((\pi(k) - \pi(i_1)) - (\pi'(k) - \pi'(i_1))\right) = b_1 - \alpha - \pi(i_1) - b_1 + b_1$$
$$= b_1 - \alpha - \pi(i_1)$$

$$(|\pi(k) - \pi(i_2)| - |\pi'(k) - \pi'(i_2)|) = \pi(i_2) - b_1 + \alpha - a_2 + b_1$$

$$S_\pi(G) - S_{\pi'}(G) = \alpha + b_1 - \alpha - \pi(i_1) + \pi(i_2) + \alpha - a_2$$
$$= (b_1 - \pi(i_1)) + \alpha + (\pi(i_2) - a_2) > 0$$

From assumption $\pi(k) < b_1$ follows that $\pi$ is not an optimal labeling. Similarly, we can prove that from assumption $\pi(k) > a_2$ follows that $\pi$ is not an optimal labeling. Q.e.d.

2) $b_2 \leq a_1$. Similarly to 1) let us assume $\pi(k) < b_2$; $\pi(k) = b_2 - \alpha$. Then we introduce a new labeling $\pi'$ by changing labels for three nodes: $\pi'(k) = b_2$; $\pi'(i_1) = a_1$; $\pi'(i_2) = b_2$. $S_\pi(G) - S_{\pi'}(G) > 0$ so we have a contradiction with the statement that $\pi(G)$ is an optimal labeling.

Lemma 7. Parent Optimal Interval - III. <u>An optimal interval of a parent is either an intersection of the optimal intervals of its children or an interval that lies between these intervals in case that their intersection is empty.</u>

If an optimal integer labeling $\pi$ provides the labels of two siblings $i_1$ and $i_2$ satisfying the conditions $a_1 \leq \pi(i_1) \leq b_1$ & $a_2 \leq \pi(i_2) \leq b_2$ then the label of their parent $k$ satisfies the following condition

if $([a_1, b_1] \cap [a_2, b_2]) \neq \emptyset$ then $\pi(k) \in [a_1, b_1] \cap [a_2, b_2])$ else

if $b_1 < a_2$ then $\pi(k) \in [b_1, a_2]$ else $\pi(k) \in [b_2, a_1]$.

For example, if $a_1 \leq a_2 \leq b_1 \leq b_2$ then Lemma 7 states that $\pi(k)$ satisfies the following condition $a_2 \leq \pi(k) \leq b_1$. Let us prove it similarly to as we do for Lemma 6. Suppose $\pi(k) < a_2$; $(a_2 - \pi(k)) = \alpha$. Then we introduce a new labeling $\pi'$ by changing labels for three nodes: $\pi'(k) = a_2$; $\pi'(i_1) = a_2$; $\pi'(i_2) = a_2$. $S_\pi(G) - S_{\pi'}(G) > 0$. So we have a contradiction with the statement that $\pi(G)$ is an optimal labeling.

## 4. Concluding Remarks.

The problems presented in this paper are associated with phylogenetics. The minsum problem resembles the problem of parsimonius labeling by characters. The traditional objective of a phylogenetic tree (evolutionary tree, a tree that is used to model the actual evolutionary history of a group of sequences or organisms) is to represent the evolutionary relationship between species. The contemporary species are represented by the leaves of the tree. Internal (branching) nodes represent common ancestor species that are now extinct. In modern molecular phylogeny, often the species at the leaves of a phylogenetic tree are represented by genes or stretches of genomic DNA. The problem of the parsimonial reconstruction of ancestral states for the given tree with the given states of its leaves (the most parsimonious assignment of the labels of internal nodes for a fixed tree topology) is a well studied problem [4, 9, 10].

We may define a k-tuple integer labeling **Π** of G as a mapping Π from G to a set of k-tuples composed of nonnegative integers $\Pi(v) = \{\pi_1(v), \pi_2(v), \ldots, \pi_{k(v)}(v)\}$, where

π_i(v) ≤ π_{i+1}(v) for all 1≤i<k(v). A **uniform** k-tuple integer labeling $\Pi_c$ of G is characterized by a constant k(v) for all v. The stretch of the edge *vw* in a $\Pi_c(G)$ is a simple sum $c_{vw} = \sum_{i=1}^{k} \theta|\pi_i(v) - \pi_i(w)|$. Given a uniform k-tuple integer labeling of the leaves of G the minimum sum problem is to find a labeling which minimizes the total sum of the stretches of the edges. The minimum sum problem is that of minimizing $\boldsymbol{s(G)} = \sum_{\forall\{vw\}\in E(G)} c_{vw}$ over all $\Pi_c$ for given k=c. By some modifications of the algorithms presented in this paper the minimizing k-tuple labeling can be found.

# References.

[1] F.R.K. Chung, Some problems and results in labelings of graphs, in: G. Chartland, e. al. (Eds.) The theory and applications of graphs John Willey, New York, 1981, pp. 255-263.
[2] F.R.K. Chung, On optimal linear arrangements of trees, Computer and Math. With Applications 10 (1984) 43-60.
[3] F.R.K. Chung, On the cutwidth and the topological bandwidth of a tree, SIAM J. Algebraic Discrete Methods, 6 (2) (1985) 268-277.
[4] W.M. Fitch, Towards defining the course of evolution: Minimum change for a specific tree topology, Syst Zool, 20 (1971) 406-416.
[5] M.K. Goldberg, I. Klipker, An Algorithm for a Minimal Placement of a Tree on a Line, Sakharth. SSR Mech. Acad. Moambe, 83 (1976) 553-556 (in Russian).
[6] M.A. Iordanskii, Minimal numberings of the vertices of trees, Soviet. Math. Dokl., 15 (1974) 1311-1315
[7] S.B. Needleman, C.D. Wunsch, A general method applicable to the search for similarities in the amino acid sequence of two proteins, J. Mol. Biol. , 48 (1970) 443–453.
[8] D. Sankoff, Matching sequences under deletion/insertion constraints, Proc. Natl. Acad. Sci. USA 69 (1972) 4-6.
[9] D. Sankoff, Minimal Mutation Trees of Sequences, SIAM J Appl Math, 28 (1975) 35-42.
[10] D. Sankoff, P. Rousseau, Locating the Vertices of a Steiner Tree in an Arbitrary Metric Space Math Program, 9 (1975) 240-246.
[11] T.F. Smith, M.S. Waterman, Identification of Common Molecular Subsequences, J. Mol. Biol., 147 (1981) 195–197.
[12] T.K. Vintsyuk, Speech discrimination by dynamic programming, Kibernetika, 4 (1968) 81-88.